\documentclass[%
 reprint,superscriptaddress,
 amsmath,amssymb,
 aps
]{revtex4-2}
\usepackage{graphicx}
\usepackage{dcolumn}
\usepackage{bm}
\usepackage[caption=false]{subfig}
\usepackage{floatrow}
\usepackage{amsmath}
\usepackage[italicdiff]{physics}
\usepackage{xcolor}
\usepackage{times}
\usepackage{amssymb}
\usepackage{physics}
\usepackage{hyperref}
\usepackage{relsize}
\DeclareCaptionLabelFormat{adja-page}{\hrulefill\\#1 #2 \emph{(previous page)}}

\date{November 3, 2023}

\begin{document}

\title{
Quantitative assessment of the universal thermopower in the Hubbard model
}
\author{Wen O. Wang}
\email{wenwang.physics@gmail.com}
\affiliation{Department of Applied Physics, Stanford University, Stanford, CA 94305, USA}
\affiliation{Stanford Institute for Materials and Energy Sciences,
SLAC National Accelerator Laboratory, 2575 Sand Hill Road, Menlo Park, CA 94025, USA}

\author{Jixun K. Ding}
\affiliation{Department of Applied Physics, Stanford University, Stanford, CA 94305, USA}
\affiliation{Stanford Institute for Materials and Energy Sciences,
SLAC National Accelerator Laboratory, 2575 Sand Hill Road, Menlo Park, CA 94025, USA}

\author{Edwin W. Huang}
\affiliation{Department of Physics and Institute of Condensed Matter Theory, University of Illinois at Urbana-Champaign, Urbana, IL 61801, USA}
\affiliation{Department of Physics and Astronomy, University of Notre Dame, Notre Dame, IN 46556, USA}
\affiliation{Stavropoulos Center for Complex Quantum Matter, University of Notre Dame, Notre Dame, IN 46556, USA}

\author{Brian Moritz}
\affiliation{Stanford Institute for Materials and Energy Sciences,
SLAC National Accelerator Laboratory, 2575 Sand Hill Road, Menlo Park, CA 94025, USA}

\author{Thomas P. Devereaux}
\email{tpd@stanford.edu}
\affiliation{Stanford Institute for Materials and Energy Sciences,
SLAC National Accelerator Laboratory, 2575 Sand Hill Road, Menlo Park, CA 94025, USA}
\affiliation{
Department of Materials Science and Engineering, Stanford University, Stanford, CA 94305, USA}
\affiliation{
Geballe Laboratory for Advanced Materials, Stanford University, Stanford, CA 94305, USA}

\begin{abstract}
As primarily an electronic observable, the room-temperature thermopower $S$ in cuprates provides possibilities for a quantitative assessment of the Hubbard model.
Using determinant quantum Monte Carlo, we demonstrate agreement between Hubbard model calculations and experimentally measured room-temperature $S$ across multiple cuprate families, both qualitatively in terms of the doping dependence and quantitatively in terms of magnitude. We observe an upturn in $S$ with decreasing temperatures,
which possesses a slope comparable to that observed experimentally in cuprates. From our calculations, the doping at which $S$ changes sign occurs in close proximity to a vanishing temperature dependence of the chemical potential at fixed density. Our results emphasize the importance of interaction effects in the systematic assessment of the thermopower $S$ in cuprates.
\end{abstract}

\maketitle

\section*{Introduction}
The Hubbard model, despite decades worth of study, remains enigmatic as a model to describe strongly correlated systems. Due to the fermion sign problem and exponential complexity, only one-dimensional systems have lent themselves to error-free estimations of ground states and their properties.
Recently, angle-resolved photoemission studies have demonstrated that a one-dimensional Hubbard-extended Holstein model can quantitatively reproduce spectra near the Fermi energy~\cite{doi:10.1126/science.abf5174,PhysRevLett.127.197003,tang2022traces}.
In two dimensions, the community lacks exact results in the thermodynamic limit; nevertheless, many of the extracted properties from simulations of the Hubbard model bear a close resemblance to observables measured in experiments, particularly those performed on high temperature superconducting cuprates.
These properties include the appearance of antiferromagnetism near half-filling, stripes, and strange metal behavior~\cite{RevModPhys.66.763,doi:10.1146/annurev-conmatphys-031620-102024,doi:10.1146/annurev-conmatphys-090921-033948}. 
However, quantitative assessments have remained out of reach, particularly regarding transport properties, where multi-particle correlation functions (calculations involving the full Kubo formalism) are computationally intensive, or one must rely on single-particle quantities (i.e. Boltzmann formalism), which can be conceptually problematic for strong interactions.

In principle, the high temperature behavior of the thermopower (thermoelectric power, or Seebeck coefficient) $S$ offers the possibility to directly test the Hubbard model against experiments in strongly correlated materials like the cuprates. Above the Debye temperature, phonons are essentially elastic scatterers of electrons and one might expect thermal relaxation to come overwhelmingly from inelastic scattering off of other electrons. Moreover, room temperature measurements afford direct contact with determinant quantum Monte Carlo (DQMC)~\cite{DQMC1,DQMC2} simulations, which are limited by the fermion sign problem to temperatures above roughly $J/2$ (half of the spin-exchange energy). Thus, one can address directly an essential question -- can the Hubbard model give both qualitative and quantitative agreement with the observed thermopower in cuprates at high temperatures?

Systematic studies of the room-temperature thermopower across a wide variety of cuprates~\cite{PhysRevB.35.8794,rao1990systematics,PhysRevB.46.14928,PhysRevB.51.12911,kaiser1995systematic,PhysRevB.59.192,PhysRevB.77.184520,PhysRevB.84.144503,PhysRevB.89.155101} show that the thermopower falls roughly on a universal curve over a broad range of hole doping $p$, with a more-or-less universal sign change near optimal doping. This sign change has been interpreted as evidence for a Lifshitz transition~\cite{PhysRevLett.73.1695,PhysRevB.54.12569,PhysRevB.84.245107}; however, this implies that the doping associated with the sign change depends on material specifics and the detailed shapes of Fermi surfaces, which is hard to reconcile with the observed universality. 
An alternative interpretation of the sign change appeals to the atomic limit~\cite{mukerjee2007doping,PhysRevB.10.2186,PhysRevB.13.647,PhysRevB.72.195109,phillips2009mottness,PhysRevB.82.214503,PhysRevLett.122.186601};
however, the atomic limit requires extremely strong interactions and a very high temperature $T$ compared to the bandwidth, neither of which is satisfied in cuprates at room temperature.
The thermopower $S$ also has been approximated by the entropy per density, defined through the Kelvin formula $S_\mathrm{Kelvin}=(\partial s/\partial n)_T/e^*$~\cite{PhysRevB.82.195105}, where charge $e^*=-e$ for electrons. $S_\mathrm{Kelvin}$ is believed to be an accurate proxy for the thermopower $S$, since it accounts for the full effects of interactions, while bypassing the difficulties in exactly calculating the Kubo formula~\cite{PhysRevB.82.195105,phillips2009mottness,Garg_2011,PhysRevB.87.035126}. However, a direct comparison between $S$ and $S_\mathrm{Kelvin}$ is required before drawing any conclusions based on these assumptions. 

Here, we calculate the thermopower $S$ based on the many-body Kubo formula, as well as the Kelvin formula $S_\mathrm{Kelvin}$, for the $t$-$t'$-$U$ Hubbard model. We employ numerically exact DQMC and maximum entropy analytic continuation (MaxEnt)~\cite{jarrell1996bayesian,PhysRevB.82.165125} to obtain the DC transport coefficients that specifically enter the evaluation of $S$. Our results show that the Hubbard model can quantitatively capture the magnitudes and the general patterns of $S$ that have been observed in cuprate experiments.

\begin{figure}
    \centering
    \includegraphics[width=\textwidth]{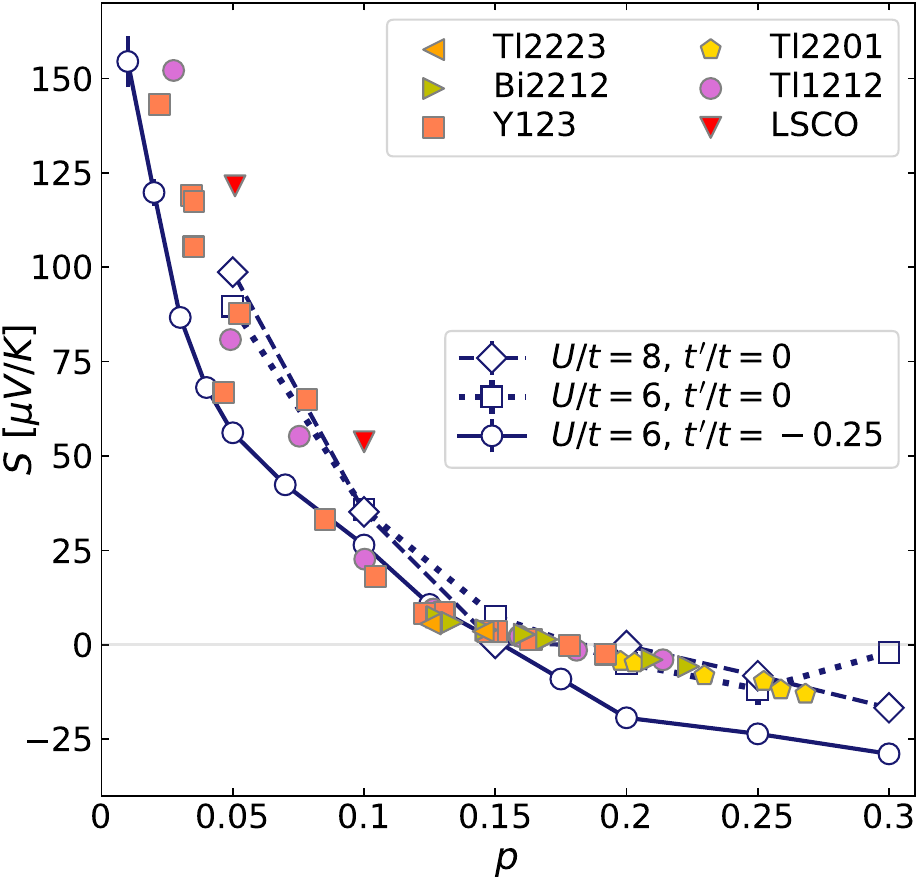}
    \caption{{\bf Comparison of simulated and experimental thermopower.} Thermopower $S$ as a function of doping $p$ from DQMC simulations (empty markers connected by lines), compared with doping dependence of $S$ for various cuprates at $T=290\, K$ (solid scattered markers, data from Refs.~\cite{PhysRevB.46.14928,PhysRevB.35.8794}).
    For $U/t=8$ and $t'/t=0$, the temperature is $k_B T=t/3.5$.
    For $U/t=6$, the temperature is $k_B T=t/4$ for both $t'/t=0$ and $t'/t=-0.25$.
}
    \label{fig:dopdep}
\end{figure}

\section*{Results}
The doping dependence of thermopower $S$ from the Hubbard model is shown in Fig.~\ref{fig:dopdep} for three different sets of parameters at their lowest achievable temperatures, overlaid with experimental data from several families of cuprates. 
It is important to note that in the process of converting our results to real units based on universal physical quantities $k_B$ and $e$, there are no adjustable parameters: $S$ is a ratio, so the standard units of $t$ (or $U$) in the Hubbard model factor out. The most striking observation is the surprisingly good agreement between our results and the room-temperature thermopower in cuprates, in both qualitative trend and quantitative magnitudes. 
Both the simulation and experimental data show a sign change roughly at $p\sim 0.15$.
In both cases, $S$ -- a quantity proportional to the electronic resistivity -- increases dramatically in the low doping regime, as the system approaches a Mott insulator.
The simulation shows moderate $U$ and $t'$ dependence,
without significantly affecting agreement with experiments.
The moderate parameter dependence is consistent with the observed approximate universality of the doping dependence of the room-temperature $S$ for different cuprates, which may have varying effective $U$ and $t'$. 

\begin{figure*}
    \centering
    \includegraphics[width=0.8\textwidth]{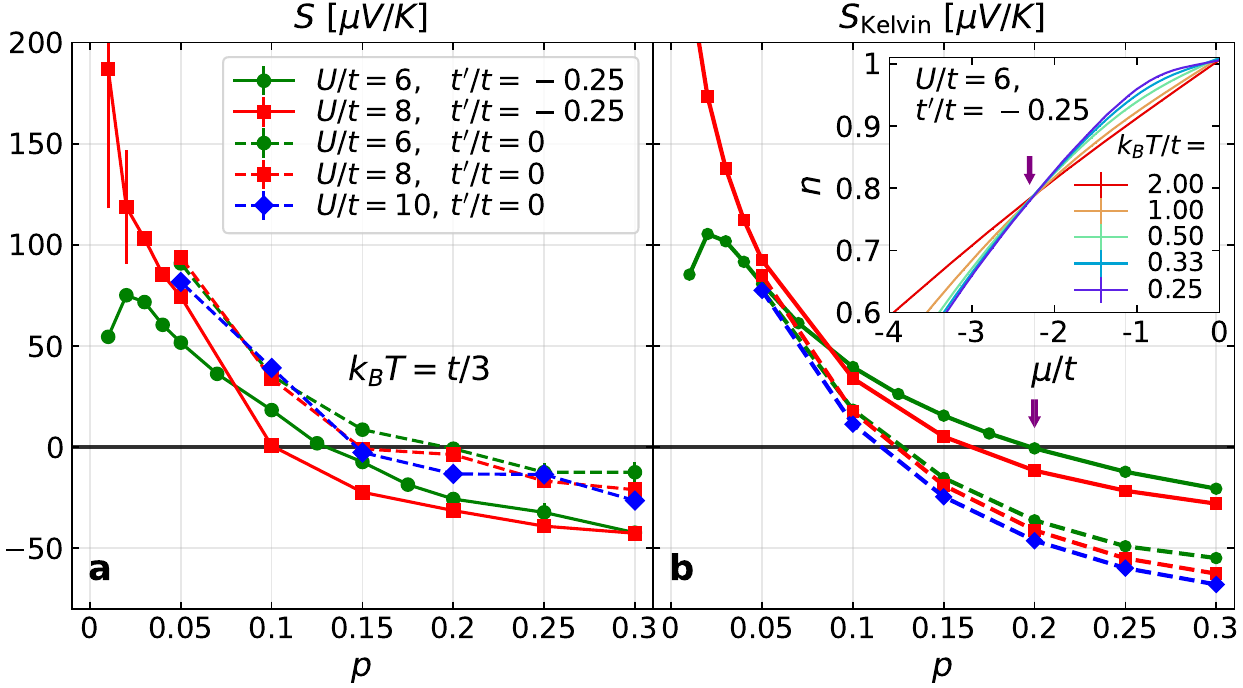}
    \caption{{\bf Doping dependence and sign change of $S$ and $S_\mathrm{Kelvin}$.} Thermopower $S$ (\textbf{a}) and the Kelvin formula for the thermopower $S_{\mathrm{Kelvin}}$ (\textbf{b}) as a function of doping $p$ for the Hubbard model with different $U$ and $t'$, all at the same temperature $k_B T=t/3$.
    Inset of (\textbf{b}): density $n$, measured using DQMC, as a function of the chemical potential $\mu$ for $U/t=6$ and $t'/t=-0.25$ at different temperatures $T$.
    The arrows in (\textbf{b}) and its inset indicate the correspondence between the sign change of $S_\mathrm{Kelvin}$ and the vanishing of temperature dependence of $\mu$ at fixed density.
}
    \label{fig:signchange}
\end{figure*}

For weakly interacting electrons, $S$ is expected to change sign around the Lifshitz transition. 
The sign change in our model with strong interactions, which occurs at $p\sim 0.15$ for $t'/t=-0.25$, is much lower than the Lifshitz transition, which occurs at $p \sim 0.26$ for the same parameters, nor is it associated with the atomic limit (see Supplementary Note~3 and Supplementary Note~5 for details).
Therefore, we seek deeper understanding from $S_\mathrm{Kelvin}= -(\partial s/\partial n)_T/e$, entropy variation per density variation at a fixed temperature, or equivalently, by the Maxwell relation, $(\partial \mu/\partial T)_n/e$, chemical potential variation per temperature variation at fixed density (see Supplementary Note~4).
In Fig.~\ref{fig:signchange}, we compare the doping dependence of $S$ and $S_\mathrm{Kelvin}$.
Despite differences in exact values, the sign change of $S$, as shown in Fig.~\ref{fig:signchange}a, is closely associated with that of $S_\mathrm{Kelvin}$, as shown in Fig.~\ref{fig:signchange}b. The sign change of $S_\mathrm{Kelvin}$ occurs when the temperature dependence of the chemical potential $\mu$ vanishes at fixed density -- an ``isosbestic'' point, as exemplified in the inset of Fig.~\ref{fig:signchange}b, and highlighted by the arrows.

The doping dependence of $S$ and $S_\mathrm{Kelvin}$ are also qualitatively similar,
and $U$ generally affects both $S$ and $S_\mathrm{Kelvin}$ in a similar manner, moderately reducing the doping at which each changes sign as $U$ increases. However, $t'$ has more significant and opposite effects on $S$ and $S_\mathrm{Kelvin}$.
Comparing Fig.~\ref{fig:signchange}a and \ref{fig:signchange}b shows us that even though $S_\mathrm{Kelvin}$, a thermodynamic quantity, differs from $S$, since it does not reflect the dynamics captured by transport~\cite{Shastry_2008},
$S_\mathrm{Kelvin}$ still reflects the most important effects from the Hubbard interaction, showing a doping dependence and sign change similar to $S$.

\begin{figure*}
    \centering
    \includegraphics[width=0.8\textwidth]{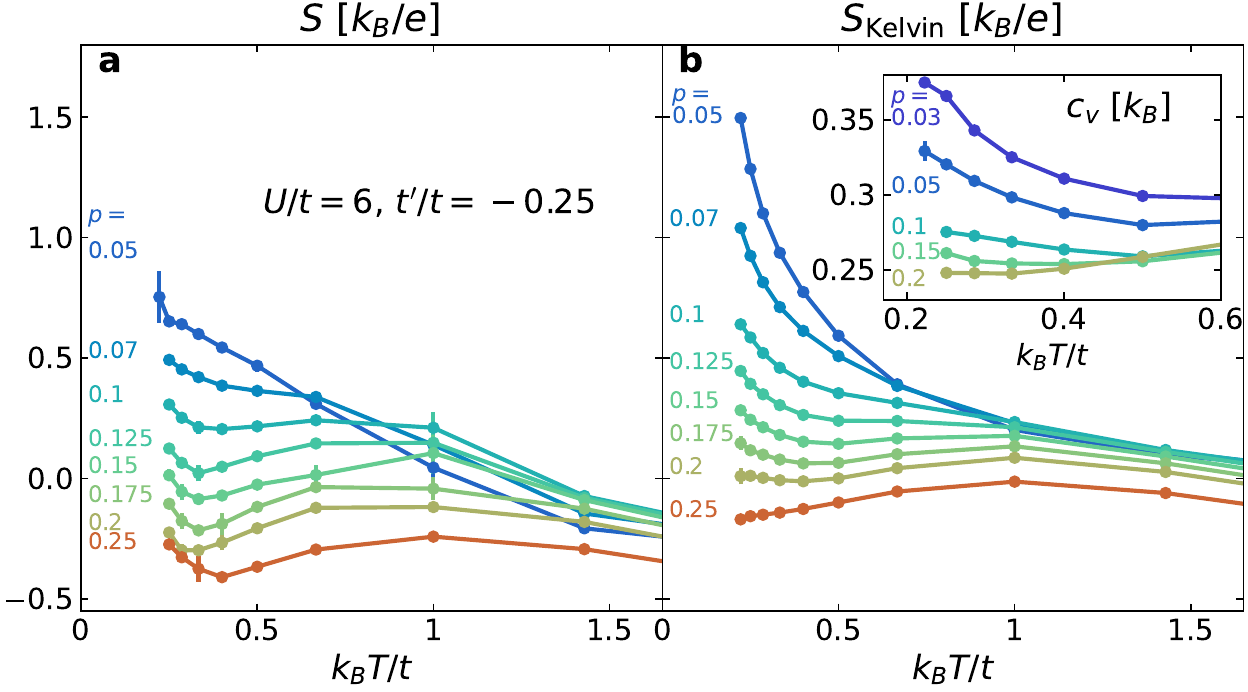}
    \caption{{\bf Temperature dependence of $S$ and $S_\mathrm{Kelvin}$.} Thermopower $S$ (\textbf{a}), and the Kelvin formula for the thermopower $S_{\mathrm{Kelvin}}$ (\textbf{b}), as a function of temperature $T$, at different doping levels $p$, for $U/t=6$ and $t'/t=-0.25$. 
    Inset of (\textbf{b}) shows the specific heat $c_v$ measured using DQMC as a function of temperature for different doping levels.
}
    \label{fig:TdepU6tp025}
\end{figure*}

We now examine the temperature dependence of $S$ and $S_{\mathrm{Kelvin}}$, using $U/t=6$ and $t'/t=-0.25$, shown in Fig.~\ref{fig:TdepU6tp025}, as a representative example.
The temperature dependence of $S$ in Fig.~\ref{fig:TdepU6tp025}a and $S_{\mathrm{Kelvin}}$ in Fig.~\ref{fig:TdepU6tp025}b are qualitatively similar.
As temperature decreases from high temperatures, $S$ and $S_{\mathrm{Kelvin}}$ first increase, following the atomic-limit ($t, t'\ll k_B T,U$, see Supplementary Note~5).
As temperature decreases further and passes the scale $t/k_B$, their behaviors deviate from the atomic-limit.
At low doping ($p\lesssim 0.07$), $S$ and $S_\mathrm{Kelvin}$ monotonically increase, but at higher doping levels, they first decrease before increasing again down to the lowest temperature, with a dip appearing in between.

\begin{figure*}
    \centering
    \includegraphics[width=0.9\textwidth]{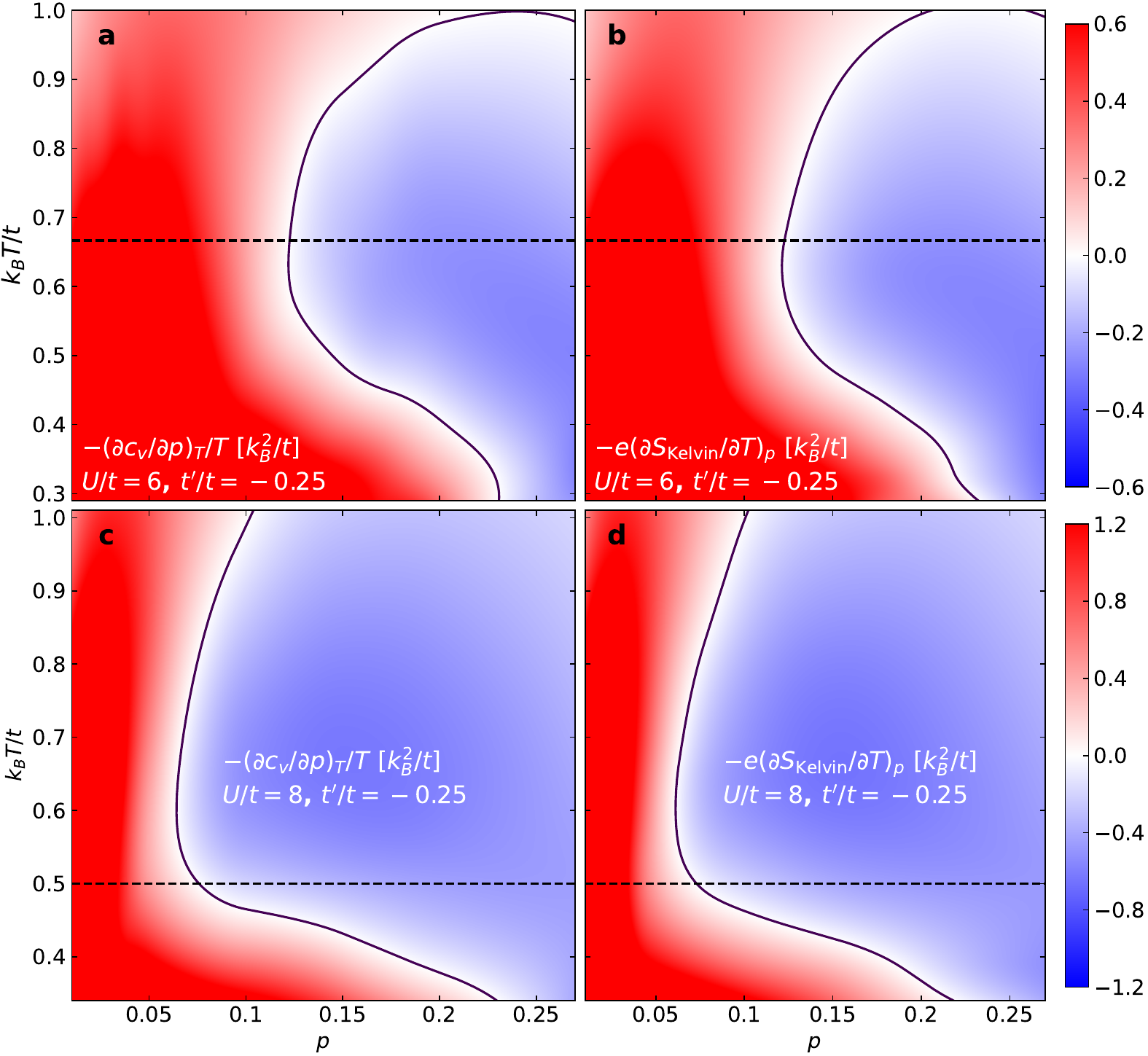}
    \caption{{\bf Analysis using $c_v$ and $S_\mathrm{Kelvin}$.} Color density plots of $-\partial^2 s/(\partial p\partial T)$ calculated from doping derivative of specific heat [$-(\partial c_v/\partial p)_T/T$, (\textbf{a} and \textbf{c})] and temperature derivative of $S_\mathrm{Kelvin}$ [$-e(\partial S_\mathrm{Kelvin}/\partial T)_p$, (\textbf{b} and \textbf{d})], for interaction strengths $U/t=6$ (\textbf{a} and \textbf{b}) and $U/t=8$ (\textbf{c} and \textbf{d}), both with $t'/t=-0.25$.
    A cubic-spline fit was applied to curves of $c_v$ versus $p$ and $S_\mathrm{Kelvin}$ versus $T$, with corresponding derivatives obtained from the fits. 
    The derivatives $-\partial^2 s/(\partial p\partial T)$ were interpolated (cubic) onto the two-dimensional $(p,T)$ plane.
    Horizontal dashed lines mark the leading-order approximation for the spin-exchange energy $J=4t^2/U$, and solid lines mark the contour where $-\partial^2 s/(\partial p\partial T)=0$.
}
    \label{fig:dsdndt}
\end{figure*}

We find the dip and the low-temperature increase in both $S$ and $S_\mathrm{Kelvin}$ particularly interesting, since this upturn commonly appears in cuprates~\cite{rao1990systematics,PhysRevB.46.14928,PhysRevB.84.144503,kaiser1995systematic,PhysRevB.35.8794}, and cannot be understood in either the atomic or weakly interacting limits.
To understand its origin, we consider the relationship between $S_\mathrm{Kelvin}$ and the specific heat $c_v$ using the Maxwell relation $-e (\partial  S_{\mathrm{Kelvin}}/\partial T)_p =-(\partial c_v/\partial p)_T/T$, where, by definition, $S_{\mathrm{Kelvin}}=( \partial s/\partial p)_T/e$ and $c_v=T(\partial s/\partial T)_p$.
Specific heat $c_v$ results, also for $U/t=6$ and $t'/t=-0.25$, are shown in the inset of Fig.~\ref{fig:TdepU6tp025}b.
Near half-filling and for temperatures below the spin-exchange energy $J$ ($=4t^2/U$ to leading order), 
$c_v$ starts to increase with decreasing temperatures, which is believed to be associated with spin fluctuations~\cite{PhysRevB.63.125116,PhysRevB.55.12918,PhysRevA.86.023633}, and $c_v$ drops with increasing doping.
Correspondingly, $S_\mathrm{Kelvin}$ at fixed doping increases with decreasing temperatures, leading to the low-temperature upturn.
As the upturn is a common feature shared by $S$ and $S_\mathrm{Kelvin}$, it is reasonable to believe that the origin should be the same.

The low-temperature slope of the thermopower can be compared with experiments. 
The negative slopes quoted in Ref.~\cite{PhysRevB.46.14928} for Bi$_2$Sr$_2$CaCu$_2$O$_{8+\delta}$ and Tl$_2$Ba$_2$CuO$_{6+\delta}$ range roughly from $-0.05$ to $-0.02 \, \mu V/K^2$. Assuming $t/k_B\sim 4000K$, this range corresponds to $\left[-2.3,-0.9\right]\, k_B^2/(te)$ in our model. We estimate the slope in our model by taking
the finite difference between temperatures $k_B T=t/4$ and $t/3.5$ in Fig.~\ref{fig:TdepU6tp025}a and \ref{fig:TdepU6tp025}b. For doping between $p=0.1$ and $0.2$, the calculated slope ranges between $\left[-2.1,-1.5\right]\, k_B^2/(te)$ for $S$, and $\left[-1.8,-0.2\right] \, k_B^2/(te)$ for $S_\mathrm{Kelvin}$.
Even though systematic and statistical errors in $S$ introduce uncertainties to this slope estimate, the ranges are roughly comparable between simulated $S$, $S_\mathrm{Kelvin}$, and experimental values.

For a detailed verification and analysis of the relationship between $S_\mathrm{Kelvin}$ and $c_v$, we calculate $-\partial^2 s/(\partial p\partial T)$ from derivatives of independently measured $S_{\mathrm{Kelvin}}$ and $c_v$, for both $U/t=6$ and $U/t=8$ with $t'/t=-0.25$, as shown in Fig.~\ref{fig:dsdndt}. 
Results from the two methods are consistent, up to minor discrepancies such as taking derivatives from discrete data points.
At any point along the contour $\partial^2 s/(\partial p\partial T)=0$ (black solid lines), either a peak or a dip will occur in $S_\mathrm{Kelvin}$ as a function of $T$.
We observe that a peak appears at temperatures above $J/k_B$ (dashed horizontal line) and a dip appears at temperatures below $J/k_B$.
Note that $T\sim J/k_B$ corresponds roughly to the crossover between a peak or dip in $S_{\mathrm{Kelvin}}$ for both $U/t=6$ and $U/t=8$ (c.f.~Supplementary Fig.~6), supporting our idea that the non-monotonic temperature dependence of both $S_\mathrm{Kelvin}$ and $S$ should be associated with effects of spin exchange.

\section*{Discussion}
In summary, we calculated the thermopower $S$ and the Kelvin formula $S_\mathrm{Kelvin}$ in the Hubbard model.
$S$ shows qualitative and quantitative agreement with the universal curve of the room-temperature $S$ in cuprates, with a sign change corresponding to an ``isosbestic'' point in $n$ versus $\mu$. 
$S$ and $S_\mathrm{Kelvin}$ show qualitatively similar doping dependence, and the doping at which $S$ changes sign corresponds well to that of $S_\mathrm{Kelvin}$.
As a function of temperature, we observe a low-temperature upturn in $S$ and $S_\mathrm{Kelvin}$ with a slope quantitatively comparable with the corresponding linear increase in cuprates, and we provide evidence supporting their association with the scale of $J$.
With this general agreement, we demonstrate that major features in the universal behavior of $S$ in cuprates can be replicated through a quantitative assessment of $S$ in the Hubbard model.
The observation that $S_\mathrm{Kelvin}$ captures qualitative features of $S$ enables us to understand the experimental thermopower results from the perspective of entropy variation with density.  

We emphasize the significance in such a high level of agreement between simulations and experiments for thermopower. Transport properties can be sensitive to numerous factors, which may be different between cuprates and the $t$-$t'$-$U$ Hubbard model. The combination of the model's simple form and capability to reproduce universal features suggests the dominance of interaction effects in the origin of the systematic behavior in the cuprates. Our observations highlight the importance of pursuing high-accuracy simulations accounting for the full effect of interactions in making progress at understanding these enigmatic materials.

\section*{Methods}

We investigate the two-dimensional single-band $t$-$t'$-$U$ Hubbard model with spin $S=1/2$ on a square lattice using determinant quantum Monte Carlo (DQMC)~\cite{DQMC1,DQMC2}.
The Hamiltonian is
\begin{align}
H &= - t \sum\limits_{\langle lm \rangle, \sigma}\left(c^\dagger_{l,\sigma}c_{m,\sigma}
 + \mathrm{h.c.}\right)  \nonumber \\
  &- t' \sum\limits_{\langle \langle lm \rangle \rangle, \sigma} \left(c^\dagger_{l,\sigma}c_{m,\sigma}
 + \mathrm{h.c.}\right)   \nonumber \\
 &+ U\sum\limits_{l} \left(n_{l,\uparrow}-\frac{1}{2}\right)\left(n_{l,\downarrow}-\frac{1}{2}\right),
\label{hubbard}
\end{align}
where $t$ ($t'$) is the nearest-neighbour (next-nearest-neighbour) hopping, $U$ is the on-site Coulomb interaction,
$\mathit{c}_{l,\mathit{\sigma}}^{\dagger}$ $(\mathit{c}_{l,\mathit{\sigma}})$ is the creation (annihilation) operator for an electron at site $l$ with spin $\mathit{\sigma}$, and $\mathit{n}_{l,\mathit{\sigma}} \equiv \mathit{c}_{l,\mathit{\sigma}}^{\dagger} \mathit{c}_{l,\mathit{\sigma}}$ is the number operator at site $l$ with spin $\mathit{\sigma}$. 

The Kelvin formula for the thermopower $S_\mathrm{Kelvin}$ can be calculated using DQMC through
\begin{align}
    S_\mathrm{Kelvin} = -\frac{\ev{(H-\mu N)N}-\ev{H-\mu N}\ev{N}}{eT(\ev{NN}-\ev{N}\ev{N})}, \label{skelvinexpr}
\end{align}
where $N=\sum_l (n_{l,\uparrow} + n_{l,\downarrow})$ is the total electron number operator, and $\mu$ is the chemical potential.

From the Hamiltonian in Eq.~\eqref{hubbard}, the particle current $\mathbf{J}$ and the energy current $\mathbf{J}_E$ are obtained as~\cite{wang2022wiedemann,PhysRevB.105.L161103}
\begin{align}
 \mathbf{J}
 &=\frac{t}{2}\sum\limits_{l,\bm{\delta} \in \mathrm{NN},\sigma}\bm{\delta}
\left(ic^\dagger_{l+\delta,\sigma}c_{l,\sigma}+\mathrm{h.c.}\right)  \nonumber \\
&+\frac{t'}{2}\sum\limits_{l,\bm{\delta}' \in \mathrm{NNN},\sigma}\bm{\delta}'
\left(ic^\dagger_{l+\delta ',\sigma}c_{l,\sigma}+\mathrm{h.c.}\right) \label{j}
\end{align}
and
\begin{align}
\mathbf{J}_E 
&= \sum\limits_{\substack{l,\bm{\delta}_1 \in \mathrm{NN},\\
\bm{\delta}_2 \in \mathrm{NN},\sigma}}\left(-\frac{\bm{\delta}_1+\bm{\delta}_2}{4}\right) {t^2}\left(i c^{\dagger}_{l+\delta_1+\delta_2,\sigma}c_{l,\sigma} + \mathrm{h.c.}\right) \nonumber\\
&+ \sum\limits_{\substack{l,\bm{\delta}\in \mathrm{NN},\\ \bm{\delta} ' \in \mathrm{NNN},\sigma }}\left(-\frac{\bm{\delta}+\bm{\delta} '}{2}\right)tt'\left(i c^{\dagger}_{l+\delta+\delta ',\sigma}c_{l,\sigma} + \mathrm{h.c.}\right) \nonumber\\
&+ \sum\limits_{\substack{l,\bm{\delta}'_1 \in \mathrm{NNN},\\ \bm{\delta}'_2 \in \mathrm{NNN},\sigma}}\left(-\frac{\bm{\delta}'_1+\bm{\delta}'_2}{4}\right) {t'^2} \left(i c^{\dagger}_{l+\delta '_1+\delta '_2,\sigma}c_{l,\sigma} + \mathrm{h.c.}\right) \nonumber\\
&+\frac{Ut}{4} \sum\limits_{l,\bm{\delta} \in \mathrm{NN},\sigma} \bm{\delta} \left(n_{l+\delta,-\sigma}+n_{l,-\sigma}\right)\left(ic_{l+\delta,\sigma}^\dagger c_{l,\sigma}+\mathrm{h.c.}\right) \nonumber\\
&+\frac{Ut'}{4}\sum_{\substack{l,\sigma,\\ \bm{\delta} ' \in \mathrm{NNN}}} \bm{\delta} ' \left(n_{l+\delta ',-\sigma}+n_{l,-\sigma}\right)\left(ic_{l+\delta ',\sigma}^\dagger c_{l,\sigma}+\mathrm{h.c.}\right) \nonumber\\
&- \frac{Ut}{4}\sum\limits_{l,\bm{\delta} \in \mathrm{NN},\sigma}\bm{\delta}
\left(ic^\dagger_{l+\delta,\sigma}c_{l,\sigma}+\mathrm{h.c.}\right) \nonumber\\
&-\frac{Ut'}{4}\sum\limits_{l,\bm{\delta} '\in \mathrm{NNN},\sigma}\bm{\delta} '
\left(ic^\dagger_{l+\delta ',\sigma}c_{l,\sigma}+\mathrm{h.c.}\right). \label{je}
\end{align}
To make the notations above clear, NN (NNN) denotes the set of nearest-neighbour (next-nearest-neighbour) position displacements. Specifically, on the two-dimensional square lattice, $\mathrm{NN}=\{+\mathbf{x}, -\mathbf{x}, +\mathbf{y}, -\mathbf{y}\}$ and $\mathrm{NNN}=\{+\mathbf{x}+\mathbf{y}, -\mathbf{x}+\mathbf{y}, +\mathbf{x}-\mathbf{y}, -\mathbf{x}-\mathbf{y}\}$, where the lattice constant is set to $1$ and $\mathbf{x}$ and $\mathbf{y}$ are unit vectors.
Here, if $l$ is an arbitrary site label associated with the position vector $x_l \mathbf{x} + y_l \mathbf{y}$, and $\bm{\nu}$ is a vector adding up arbitrary elements in $\mathrm{NN}$ and $\mathrm{NNN}$, the notation $l+\nu$ represents a unique site label associated with the position $x_l \mathbf{x} + y_l \mathbf{y} + \bm{\nu}$.
The heat current is $\mathbf{J}_Q= \mathbf{J}_E-\mu \mathbf{J}$.

We calculate the thermopower
\begin{equation}
    S = -\frac{L_{J_{Q,x}J_x}}{eTL_{J_xJ_x}} \label{S_expr}
\end{equation}
using DQMC and maximum entropy analytic continuation (MaxEnt)~\cite{jarrell1996bayesian,PhysRevB.82.165125}.
Here, $J_{Q,x}$ and $J_x$ are the $x$-components of the heat current operator $\mathbf{J}_Q$ and particle current operator $\mathbf{J}$, respectively.
For arbitrary Hermitian operators $O_1$ and $O_2$, the DC transport coefficient $L_{O_1O_2} \equiv \left. L_{O_1O_2}(\omega)\right|_{\omega=0}$, where $L_{O_1O_2}(\omega)$ is determined using the Kubo formula
\begin{multline}
     L_{O_1O_2}(\omega) = \\
     \frac{1}{N_xN_y \beta} \int_0^\infty d t e^{i(\omega+i0^+) t}\int_0^\beta d\tau\langle{O_1(t-i\tau)O_2(0)}\rangle,\label{definitioncoefficient}
\end{multline}
where $t$ is real time, without confusion with the hopping matrix elements in the Hamiltonian. Here, $N_x$, $N_y$ are the sizes of the lattice along the $x$ and $y$ directions, respectively, $\beta\equiv (k_B T)^{-1}$, and 
\begin{equation}
O_1(t-i\tau) = e^{i(H-\mu N)(t-i\tau)} O_1 e^{-i(H-\mu N)(t-i\tau)}.
\end{equation}

Detailed derivations for Eqs.~\eqref{S_expr} and \eqref{skelvinexpr} are in Supplementary Note~2 and Supplementary Note~4, respectively.
For our calculation, the units for both $S$ and $S_\mathrm{Kelvin}$ are $k_B/e \approx 86.17\, \mu V/K$.

\section*{Data availability}
The data needed to reproduce the figures can be found at \url{https://doi.org/10.5281/zenodo.8286640}.

\section*{Code availability}
The source code and analysis routines can be found at \url{https://doi.org/10.5281/zenodo.8286636}.

\bibliography{main}

\section*{Acknowledgments}
We acknowledge helpful discussions with D.~Belitz, R.~L.~Greene, S.~A.~Kivelson, S.~Raghu, B.~S.~Shastry, R.~Scalettar, and J.~Zaanen.
This work at Stanford and SLAC (WOW, JKD, BM, TPD) was supported by the U.S. Department of Energy (DOE), Office of Basic Energy Sciences,
Division of Materials Sciences and Engineering. 
EWH was supported by the Gordon and Betty Moore Foundation EPiQS Initiative through the grants GBMF 4305 and GBMF 8691.
Computational work was performed on the Sherlock cluster at Stanford University and on resources of the National Energy Research Scientific Computing Center, supported by the U.S. DOE, Office of Science, under Contract no. DE-AC02-05CH11231.

\section*{Author contributions}
WOW conceived the study, performed numerical simulations, conducted data analysis, interpreted the data, and wrote the manuscript.
JKD, EWH, BM, and TPD assisted in data interpretation and contributed to writing the manuscript. 

\section*{Competing interests}
The authors declare no competing interest.

\clearpage

\section*{Supplementary Information}

\setcounter{section}{0}
\setcounter{figure}{0}
\setcounter{equation}{0}

\renewcommand{\thesection}{Supplementary Note \arabic{section}:}
\makeatletter
\renewcommand{\fnum@figure}{Supplementary Fig. \thefigure}
\makeatother

\section{Simulation parameters} \label{simulationparameters}

\begin{figure*}
    \centering
    \includegraphics[width=0.8\textwidth]{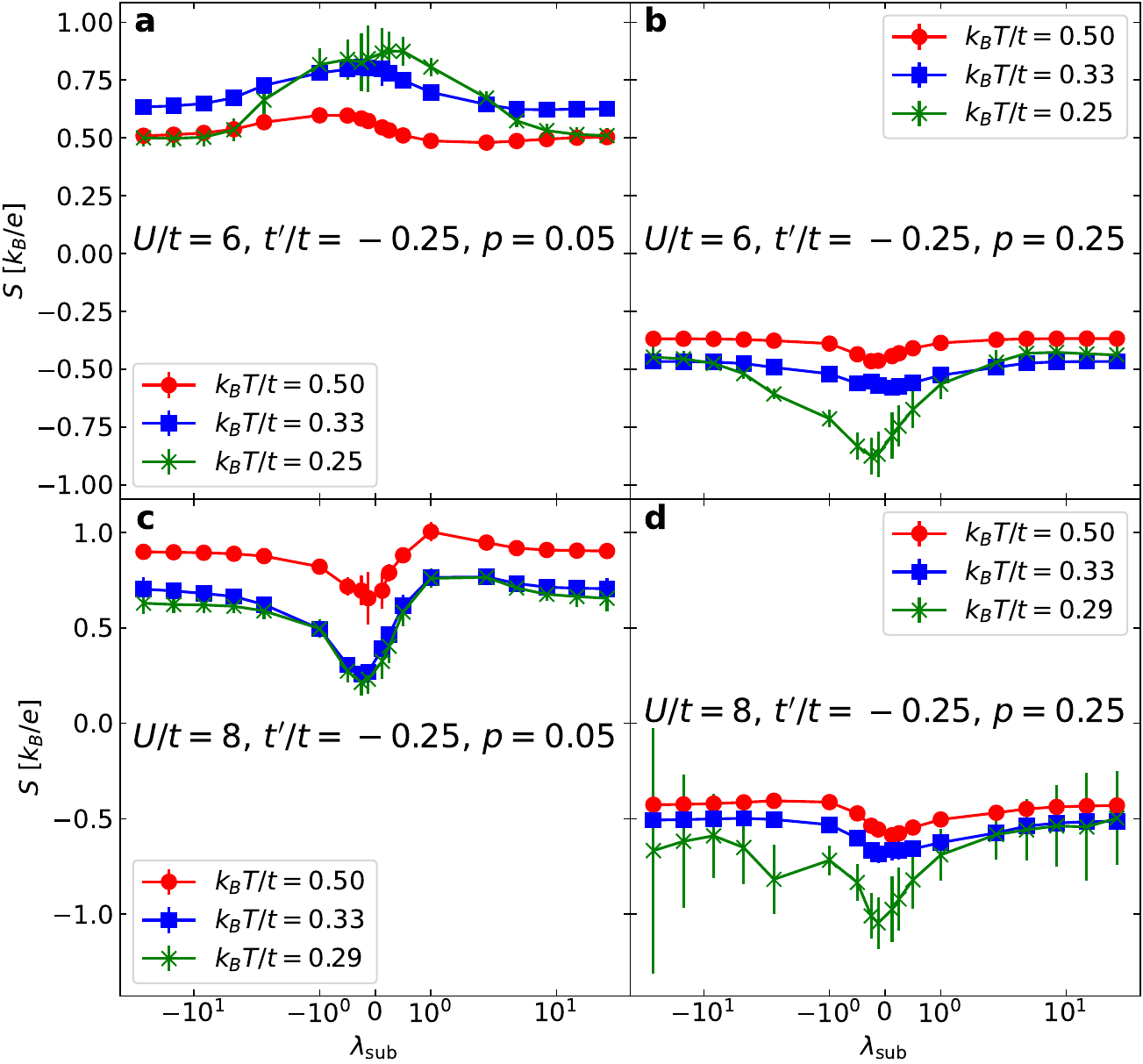}
    \caption{
$S$ as a function of $\lambda_\mathrm{sub}$ at a few different temperatures, for four sets of representative parameters, all with $t'/t=-0.25$: (\textbf{a}) $U/t = 6$, $p = 0.05$, (\textbf{b}) $U/t = 6$, $p = 0.25$, (\textbf{c}) $U/t = 8$, $p = 0.05$, and (\textbf{d}) $U/t = 8$, $p = 0.25$. For MaxEnt analytic continuation, a flat model function is used for all parameters.
 }
    \label{fig:lambdadependence}
\end{figure*}

Statistical error bars denoting $\pm 1$ standard error of the mean are shown for all measurements, except for Supplementary Fig.~\ref{fig:lifshitz} that has none.
Error bars are determined by bootstrap resampling ($100$ bootstraps)~\cite{bootstrap}, except for error bars determined by jackknife resampling~\cite{jackknife}: $n$ in the inset of Fig.~2b in the main text, $S_{\mathrm{Kelvin}}$ in Supplementary Fig.~\ref{fig:atomiclimit_largeU},
and $16\times 16$  $S_\mathrm{Kelvin}$ data in Supplementary Fig.~\ref{fig:finitesizeandtrotter}b.
Simulation cluster size is $8\times 8$ for all results, unless otherwise specified.
The maximum imaginary time Trotter discretization is $d\tau=0.02/t$ in the chemical potential tuning process, and $d\tau=0.05/t$ for other thermodynamic and transport measurements, unless otherwise specified. 
At high temperatures, the smallest number of imaginary-time slices used in the Trotter decomposition is $\tilde{L}=\beta/d\tau = 20$.
For MaxEnt analytic continuation, we choose the model function by using the same high-temperature annealing procedure as in Ref.~\cite{wang2022wiedemann}, except for Supplementary Fig.~\ref{fig:lambdadependence}.
We determine spectra in the infinite-temperature-limit, using a moments expansion method, which serves as the model function at the highest temperature, except for Supplementary Fig.~\ref{fig:lifshitz}, similar as in Refs.~\cite{edwin,PhysRevB.105.L161103,wang2022wiedemann}.
To determine the adjustable parameter which assigns weights of statistics and entropy in the maximized function in MaxEnt, we use the method of Ref.~\cite{PhysRevE.94.023303}.
Other details in methods and parameter choices are mostly the same as Ref.~\cite{wang2022wiedemann}.

\section{Formalism}
\label{sec:formalism}
We set $\hbar$ to $1$ throughout the paper.
We consider the response due to a temperature gradient $\nabla T$ and electric field $\mathbf{E}=-\nabla V$.
We define $\overline{\mu} = \mu+e^*V$ so that $\nabla \overline{\mu} = \nabla \mu - e^* \mathbf{E}$, where charge $e^*=-e$ for electrons.
The responses along the $x$ direction in terms of DC transport coefficients $L_{O_1O_2}$ ($\omega=0$ value of Eq.~(6) in the main text) are~\cite{Shastry_2008,PhysRevB.105.L161103}
\begin{align}
& \langle J_x \rangle/(N_xN_y) = -\beta L_{J_xJ_x} \partial_{x} \overline{\mu} + L_{J_xJ_{Q,x}}\partial_{x} \beta \nonumber \\
&=-\beta L_{J_xJ_x} \partial_{x} \overline{\mu} - L_{J_xJ_{Q,x}}\beta^2 k_B\partial_{x} T,
\label{eq:current} \\
&\langle J_{Q,x} \rangle/(N_xN_y) =- \beta L_{J_{Q,x}J_x} \partial_{x} \overline{\mu} + L_{J_{Q,x}J_{Q,x}}\partial_{x} \beta \nonumber \\
&=-\beta L_{J_{Q,x}J_x} \partial_{x} \overline{\mu} - L_{J_{Q,x}J_{Q,x}}\beta^2 k_B\partial_{x} T. \label{eq:heatcurrent} 
\end{align}

The thermopower $S$ is defined as
\begin{equation}
    S = -\left.\frac{\partial_x\overline{\mu}}{e^*\partial_xT }\right|_{\ev{J_{x}}=0} = - \frac{L_{J_xJ_{Q,x}}}{eTL_{J_xJ_x}}
    = -\frac{ L_{J_{Q,x}J_x}}{eTL_{J_xJ_x}}  , \label{thermopower}
\end{equation}
giving us Eq.~(5) in the main text.
In Supplementary Eq.~\eqref{thermopower}, we used Onsager's reciprocity relations~\cite{bulusu2008review}
\begin{align}
    L_{J_xJ_{Q,x}} = L_{J_{Q,x}J_x}. \label{onsager}
\end{align}

Setting $Z=\mathrm{Tr}(e^{-\beta(H-\mu N)})$ as the partition function, from Eq.~(6) in the main text,
\begin{align}
     L_{O_1O_2}(\omega) &= \frac{1}{ZN_xN_y\beta} \sum\limits_{i_1,i_2}\langle {i_1|O_1|i_2}\rangle\langle {i_2|O_2|i_1}\rangle\nonumber \\
     &\quad \times \frac{e^{-\beta E_{i_1}}-e^{-\beta E_{i_2}}}{i(E_{i_1}-E_{i_2})(\omega+i0^++E_{i_1}-E_{i_2})}.
     \label{long}
\end{align}

In the case of $O_1=O_2=O$,
we obtain
\begin{align}
     \Re L_{OO}(\omega) &= \frac{\pi}{ZN_xN_y\beta\omega} \sum\limits_{i_1,i_2}|\langle {i_1|O|i_2}\rangle|^2\nonumber \\
     &\quad \times e^{-\beta E_{i_1}}(1-e^{-\beta\omega}) \delta(\omega+E_{i_1}-E_{i_2}),
     \label{definitioncoefficient_useful}
\end{align}
where $|i_i\rangle$ ($E_{i_1}$) are eigenstates (eigenvalues) of the grand-canonical Hamiltonian $H-\mu N$.
From Supplementary Eq.~\eqref{definitioncoefficient_useful} we obtain $\Re L_{OO}(\omega)=\Re L_{OO}(-\omega)$. By Kramers-Kronig relations, $\Im L_{OO}(\omega=0)=0$.

We use DQMC to measure correlation functions in imaginary time,
\begin{align}
&\langle{T_\tau O_1(\tau)O_2(0)}\rangle \nonumber\\
&\equiv \frac{1}{Z}\mathrm{Tr}\left(e^{-(\beta-\tau)(H-\mu N)}O_1 e^{-\tau (H-\mu N) } O_2\right) \nonumber \\
&=\frac{1}{Z}\sum\limits_{i_1,i_2} \langle{i_1|O_1|i_2}\rangle \langle{i_2|O_2|i_1}\rangle e^{-\beta E_{i_1}}e^{\tau(E_{i_1}-E_{i_2})}. \label{imaginarytimecorr}
\end{align}
Comparing Supplementary Eqs.~\eqref{definitioncoefficient_useful} and \eqref{imaginarytimecorr}, we relate $\Re L_{OO}(\omega)$ with $\langle{T_\tau O(\tau)O(0)}\rangle$ through
\begin{equation}
\frac{\langle{T_\tau O(\tau)O(0)}\rangle}{N_xN_y\beta}=  \int_0^{\infty} d\omega \Re L_{OO}(\omega) \frac{\omega\cosh [\omega(\tau-\beta/2) ]}{\pi\sinh [\beta\omega/2 ]}. \label{kernel}
\end{equation}
We apply MaxEnt analytic continuation to $\langle{T_\tau O(\tau)O(0)}\rangle$ data to invert Supplementary Eq.~\eqref{kernel} and obtain $\Re L_{OO}(\omega)$.

According to Eq.~(6) in the main text, we may write
\begin{multline}
\frac{1}{2}(L_{O_1O_2}(\omega)+L_{O_2O_1}(\omega))  = \left(L_{(\lambda_\mathrm{sub} O_1+O_2)(\lambda_\mathrm{sub}O_1+O_2)}(\omega)\right.  \\
\left.- \lambda_\mathrm{sub}^2 L_{O_1O_1}(\omega)   - L_{O_2O_2}(\omega) \right)/(2\lambda_\mathrm{sub}), \label{substractionO1O2}
\end{multline}
where $\lambda_\mathrm{sub}$ is an arbitrary non-zero real constant~\cite{PhysRevB.95.121104}.
With Supplementary Eq.~\eqref{substractionO1O2}, Supplementary Eq.~\eqref{kernel} can be generalized,
\begin{align}
&\frac{\langle{T_\tau O_1(\tau)O_2(0)}\rangle+\langle{T_\tau O_2(\tau)O_1(0)}\rangle}{N_xN_y\beta} \nonumber \\
&=  \int_0^{\infty} d\omega \Re \left[L_{O_1O_2}(\omega)+L_{O_2O_1}(\omega)\right] \frac{\omega\cosh [\omega(\tau-\beta/2) ]}{\pi\sinh [\beta\omega/2 ]}. \label{kernelO1O2}
\end{align}

$\Re L_{OO}(\omega)$ is guaranteed to be positive definite in Supplementary Eq.~\eqref{definitioncoefficient_useful} when $O_1=O_2=O$, in which case MaxEnt analytic continuation is applicable.
However, in calculation of the thermopower, $\Re L_{J_{Q,x}J_x}(\omega)+\Re L_{J_xJ_{Q,x}}(\omega)$ can change its sign as a function of $\omega$, so it cannot be directly calculated from $\ev{T_\tau J_{Q,x}(\tau)J_x}+\ev{T_\tau J_x(\tau)J_{Q,x}}$ using Supplementary Eq.~\eqref{kernelO1O2} through MaxEnt.
So, according to Supplementary Eqs.~\eqref{onsager} and \eqref{substractionO1O2}, we calculate  $L_{J_{Q,x}J_x}$ using
\begin{align}
&L_{J_{Q,x}J_x} = \left(L_{(\lambda_\mathrm{sub} J_{Q,x}+J_{x})(\lambda_\mathrm{sub} J_{Q,x}+J_{x})}\right.  \nonumber \\
&\left.- \lambda_\mathrm{sub}^2 L_{J_{Q,x}J_{Q,x}}   - L_{J_x J_x} \right)/(2\lambda_\mathrm{sub}). \label{substraction}
\end{align}
Since $L_{OO} \equiv \left. L_{O O}(\omega)\right|_{\omega=0}$ is real, $L_{J_{Q,x}J_x}$ is also real.
In principle, if there are no errors in every $L_{OO}$ term on the right hand side of Supplementary Eq.~\eqref{substraction}, then the result of $L_{J_{Q,x}J_x}$ from Supplementary Eq.~\eqref{substraction} is  $\lambda_\mathrm{sub}$ independent. 
However, systematic errors introduced by the analytic continuation process propagate in the calculation of Supplementary Eq.~\eqref{substraction}, which is reflected by $S$ exhibiting some degree of $\lambda_\mathrm{sub}$ dependence.
In Supplementary Fig.~\ref{fig:lambdadependence}, we show $S$ as a function of $\lambda_\mathrm{sub}$ for four sets of parameters as examples.
As long as $|\lambda_\mathrm{sub}| \gtrsim 1$, the $\lambda_\mathrm{sub}$ dependence is relatively weak. 
Therefore, as a reasonable choice, we use $\lambda_\mathrm{sub}=2$ in this work.

\section{Lifshitz transition} \label{sec:lifshitz}
\begin{figure}
    \centering
    \includegraphics[width=0.8\textwidth]{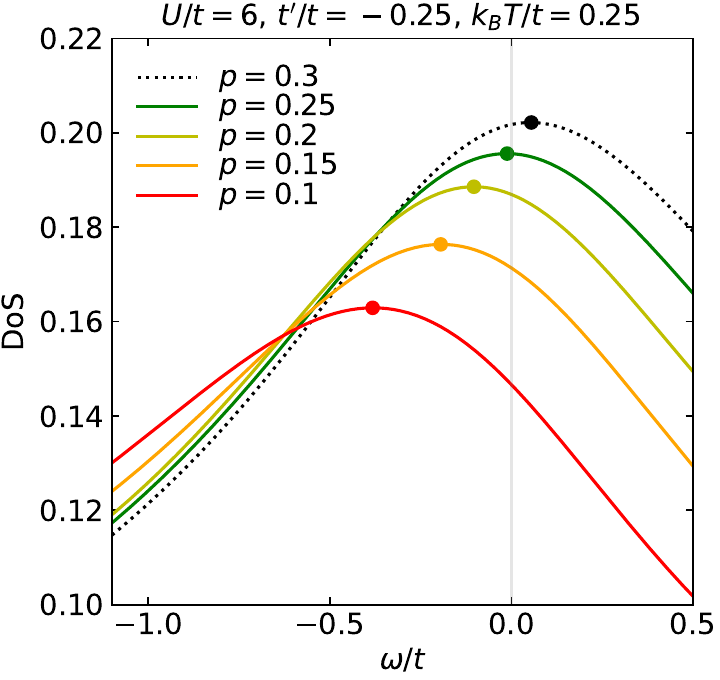}
    \caption{
Density of states (DoS) as a function of $\omega$ for different dopings, for $U/t=6$ and $t'/t=-0.25$ at $k_B T/t=0.25$. $\omega=0$ (light grey vertical line) sets the Fermi level.
 }
    \label{fig:lifshitz}
\end{figure}

We calculate the density of states (DoS) from the DQMC results of the local Green's function $G(\tau)=-\langle T_\tau c_{l,\sigma}(\tau)c_{l,\sigma}^\dagger(0) \rangle$, by inverting the relation~\cite{mahan2013many}
\begin{equation}
    G(\tau) = -\int_{-\infty}^{+\infty}d\omega\frac{e^{-\tau\omega}}{1+e^{-\beta\omega}} \mathrm{DoS}(\omega),
\end{equation}
using MaxEnt analytic continuation.
For the model function in MaxEnt, we start with using the flat model at the highest temperature $k_B T/t=8$, and proceed with lower temperatures using the high-temperature annealing procedure.
In Supplementary Fig.~\ref{fig:lifshitz} we show doping dependence of DoS$(\omega)$ for fixed $U/t=6$, $t'/t=-0.25$, and $k_B T/t=0.25$.
We observe that the Lifshitz transition, at which the quasiparticle peak crosses the Fermi level at $\omega=0$, happens at doping $p\sim 0.26$, which is much higher than the sign change doping of $S$ at $p\sim 0.15$ in Fig.~1 in the main text for the corresponding parameter set.
Therefore, the sign change doping of $S$ is not associated with the Lifshitz transition.

\begin{figure*}
    \centering
    \includegraphics[width=0.8\textwidth]{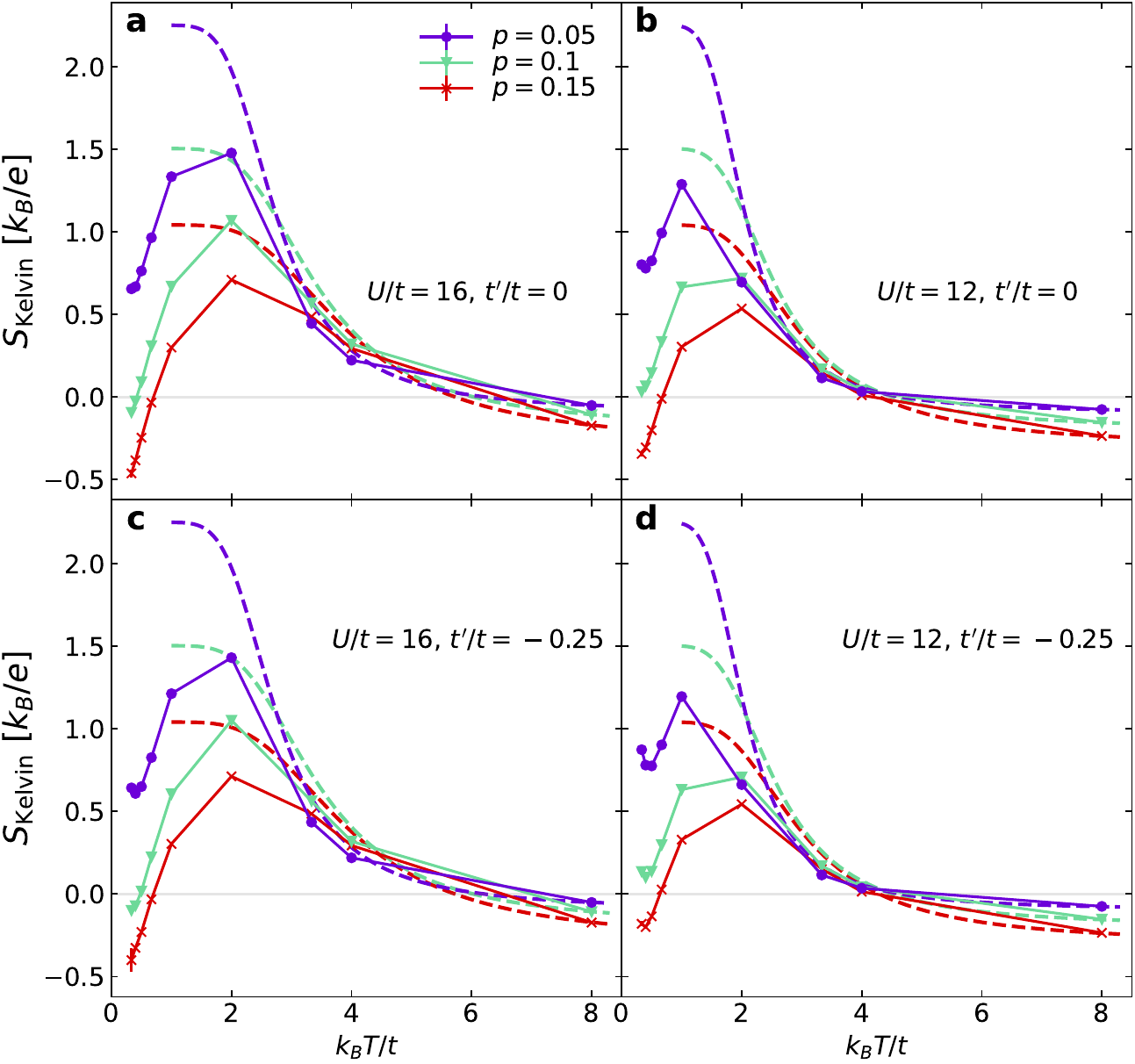}
    \caption{
    Comparison between Hubbard model (solid lines, obtained by DQMC) and atomic-limit (dashed lines, from Supplementary Eq.~\eqref{skelvin_al}) results of $S_\mathrm{Kelvin}$ for large interactions: (\textbf{a}) $U/t = 16$, $t'/t = 0$, (\textbf{b}) $U/t = 12$, $t'/t = 0$, (\textbf{c}) $U/t = 16$, $t'/t = -0.25$, and (\textbf{d}) $U/t = 12$, $t'/t = -0.25$. For these parameters,
    the maximum $d\tau$ for chemical potential tuning is $0.01/t$.
Error bars denote $\pm 1$ standard error of the mean determined by jackknife resampling.
 }
    \label{fig:atomiclimit_largeU}
\end{figure*}

\section{Kelvin formula} \label{sec:Kelvin}

The Kelvin formula for thermopower is~\cite{PhysRevB.82.195105}
\begin{align}
    S_{\mathrm{Kelvin}} = \frac{1}{e^*} \left( \frac{\partial s}{\partial n}\right)_T = -\frac{1}{e^*} \left(\frac{\partial \mu}{\partial T}\right)_n, \label{eqkelvin}
\end{align}
where $s$ is the entropy density and $n$ is the particle density.

To obtain the second equality in Supplementary Eq.~\eqref{eqkelvin}, we consider the thermodynamic potential density $f=\epsilon-sT-\mu n$, where $\epsilon$ is the energy density.
Using the first law of thermodynamics,
\begin{equation}
 \dd\epsilon = T\, \dd s +\mu \, \dd n, \label{firstlaw}
\end{equation}
we obtain $\dd (f+\mu n) = -s\,\dd T + \mu \,\dd n$. Equating 
\begin{align}
\frac{\partial^2 (f+\mu n)}{\partial T\partial n} = \frac{\partial^2 (f+\mu n)}{\partial n\partial T}
\end{align}
then gives us the Maxwell relation leading to the second equality in Supplementary Eq.~\eqref{eqkelvin}.

From Supplementary Eqs.~\eqref{eqkelvin} and \eqref{firstlaw}, we find
\begin{align}
    S_{\mathrm{Kelvin}} = \frac{1}{e^*}\left[ \frac{1}{T}\left(\frac{\partial \epsilon}{\partial n}\right)_T-\frac{\mu}{T} \right], \label{skevlin2}
\end{align}
where 
\begin{equation}
    \left(\frac{\partial \epsilon}{\partial n}\right)_T = \left(\frac{\partial \epsilon}{\partial \mu}\right)_T\bigg/ \left(\frac{\partial n}{\partial \mu}\right)_T. \label{intermediate}
\end{equation}
In terms of correlation functions, which we measure using DQMC,
\begin{align}
    \left(\frac{\partial \epsilon}{\partial \mu}\right)_T &=  \frac{\partial}{\partial \mu} \frac{\Tr H e^{-\beta (H-\mu N)}}{N_xN_y\Tr e^{-\beta (H-\mu N)}} \nonumber \\
    &=\frac{\beta}{N_xN_y}(\ev{HN}-\ev{H}\ev{N}), \label{correlator1}\\
    \left(\frac{\partial n}{\partial \mu}\right)_T &=  \frac{\partial}{\partial \mu} \frac{\Tr N e^{-\beta (H-\mu N)}}{N_xN_y\Tr e^{-\beta (H-\mu N)}} \nonumber \\
    &=\frac{\beta}{N_xN_y}(\ev{NN}-\ev{N}\ev{N}).\label{correlator2}
\end{align}
Taking Supplementary Eqs.~\eqref{skevlin2}, \eqref{intermediate}, \eqref{correlator1},  and \eqref{correlator2}, with $e^*=-e$, we obtain Eq.~(2) in the main text. 

The specific heat (considering Supplementary Eq.~\eqref{firstlaw}) is 
\begin{align}
    c_v=\left(\frac{\partial\epsilon}{\partial T}\right)_n = T\left(\frac{\partial s}{\partial T}\right)_n . \label{cvwiths}
\end{align}
So from Supplementary Eqs.~\eqref{eqkelvin} and \eqref{cvwiths}, we obtain
\begin{align}
   e^* \left(\frac{\partial  S_{\mathrm{Kelvin}}}{\partial T}\right)_n = \frac{\partial^2 s}{\partial n\partial T}= \frac{\partial^2 s}{\partial T\partial n} =\frac{1}{T}\left(\frac{\partial c_v}{\partial n}\right)_T. \label{eq0:dsdndt}
\end{align}
Therefore, the temperature dependence of $S_{\mathrm{Kelvin}}$ is directly related to doping dependence of the specific heat $c_v$.
In the main text, we use doping $p=1-n$ instead of $n$. So we rewrite Supplementary Eq.~\eqref{eq0:dsdndt} as
\begin{align}
   -e \left(\frac{\partial  S_{\mathrm{Kelvin}}}{\partial T}\right)_p = -\frac{\partial^2 s}{\partial p\partial T} = -\frac{\partial^2 s}{\partial T\partial p} =-\frac{1}{T}\left(\frac{\partial c_v}{\partial p}\right)_T. \label{eq1:dsdndt}
\end{align}

For the calculation of $-\partial^2 s/(\partial p\partial T)$ in Fig.~4 in the main text, to rule out data points with large error bars in the spline fitting process, for the fitting of $c_v$, the lowest temperature considered is $k_B T = t/3.5$ for $U/t=6$ and $k_B T = t/3$ for $U/t=8$;
for $S_\mathrm{Kelvin}$, the lowest temperature in the fitting range is $k_B T = t/4.5$ for $U/t=6$ and $k_B T = t/3.5$ for $U/t=8$.
Since the measurements of $c_v$ involve energy fluctuation and therefore contains correlators with up to $8$ fermion operators, while $S_\mathrm{Kelvin}$ contains up to $6$, for the same set of parameters, $c_v$ data generally has larger statistical error than $S_\mathrm{Kelvin}$.
Therefore a higher lowest temperature is chosen for fitting $c_v$ than that for $S_\mathrm{Kelvin}$.

\section{Atomic limit}
\label{sec:atomic-limit}
In this note we derive the atomic-limit ($t, t'\ll k_B T, U$) approximation of $S$ and $S_\mathrm{Kelvin}$.

Considering the condition $t,t'\ll U$, we divide the Hamiltonian of Eq.~(1) in the main text into the interaction part $H_0 \propto U$ as the unperturbed Hamiltonian and the kinetic part $\Delta H$ as the perturbative term. Namely,
\begin{align}
&H_0= U\sum\limits_{l} \left(n_{l,\uparrow}-\frac{1}{2}\right)\left(n_{l,\downarrow}-\frac{1}{2}\right),\nonumber \\
&\Delta H = - t \sum\limits_{\langle lm \rangle ,\sigma}\left(c^\dagger_{l,\sigma}c_{m,\sigma}
 + \mathrm{h.c.}\right)  \nonumber\\
 &\qquad \ \ \ - t' \sum\limits_{\langle \langle lm \rangle \rangle , \sigma} \left(c^\dagger_{l,\sigma}c_{m,\sigma}
 + \mathrm{h.c.}\right).  \nonumber
\end{align}
By expanding 
\begin{align}
&e^{-\tau (H-\mu N)}=e^{-\tau (H_0-\mu N)}\left[1-\int_0^\tau d\tau_1 \Delta H(\tau_1)\right. \nonumber \\
&+\left.\int_0^\tau d\tau_1 \int_0^{\tau_1}d\tau_2 \Delta H(\tau_1) \Delta H(\tau_2)+ ... \right], \label{expandd}
\end{align}
where $\Delta H(\tau_1)=e^{\tau_1 (H_0-\mu N)} \Delta H e^{-\tau_1 (H_0-\mu N)}$,
the $O_1-O_2$ correlation function between arbitrary Hermitian operators $O_1,O_2$ is
\begin{align}
    &\ev{T_\tau O_1(\tau) O_2} \nonumber \\
    &= \frac{\Tr \left(e^{-(\beta-\tau) (H -\mu N)} O_1 e^{-\tau (H-\mu N)} O_2\right)}{\Tr e^{-\beta (H-\mu N)}} \nonumber \\
    &=\frac{\Tr (e^{-(\beta-\tau) (H_0 -\mu N)}O_1 e^{-\tau (H_0-\mu N)} O_2)}{\Tr e^{-\beta (H_0 -\mu N)}}(1+\mathcal{O}(\beta t)). ~\label{correlator_AL}
\end{align}
Using Supplementary Eq.~\eqref{correlator_AL} evaluated under the occupation basis (the eigenstates of $H_0$), Eq.~(2) in the main text can be obtained to leading order. This leads to the atomic-limit approximation 
\begin{equation}
S_\mathrm{Kelvin} = \frac{-U\left(e^{2\beta\mu+\frac{\beta U}{2}}+e^{\beta \mu}\right)}{eT\left(e^{\frac{\beta U}{2}}+e^{2\beta\mu+\frac{\beta U}{2}}+2e^{\beta\mu}\right)} +\frac{\frac{U}{2}+\mu}{eT} .
\label{skelvin_al}
\end{equation}
In the same limit, we can calculate the average density $\ev{n}$. Applying Supplementary Eq.~\eqref{expandd}, we find
\begin{align}
&\ev{n} = \frac{\Tr (e^{-\beta (H-\mu N)} N)}{N_xN_y\Tr (e^{-\beta (H-\mu N)} )}\nonumber \\
& = \frac{\Tr (e^{-\beta (H_0-\mu N)} N)}{N_xN_y\Tr (e^{-\beta (H_0-\mu N)} )} (1+\mathcal{O}(\beta t)). \label{napprox}
\end{align}
Therefore, to leading order,
\begin{align}
&\ev{n} = \frac{2e^{\frac{1}{2}\beta U+\beta\mu}+2e^{2\beta\mu}}{1+2e^{\frac{1}{2}\beta U+\beta\mu}+e^{2\beta\mu}}, \label{density_al}
\end{align}
which allows us to determine $\mu$ for any given density $n$ in the atomic limit.

\begin{figure*}
    \centering
    \includegraphics[width=0.8\textwidth]{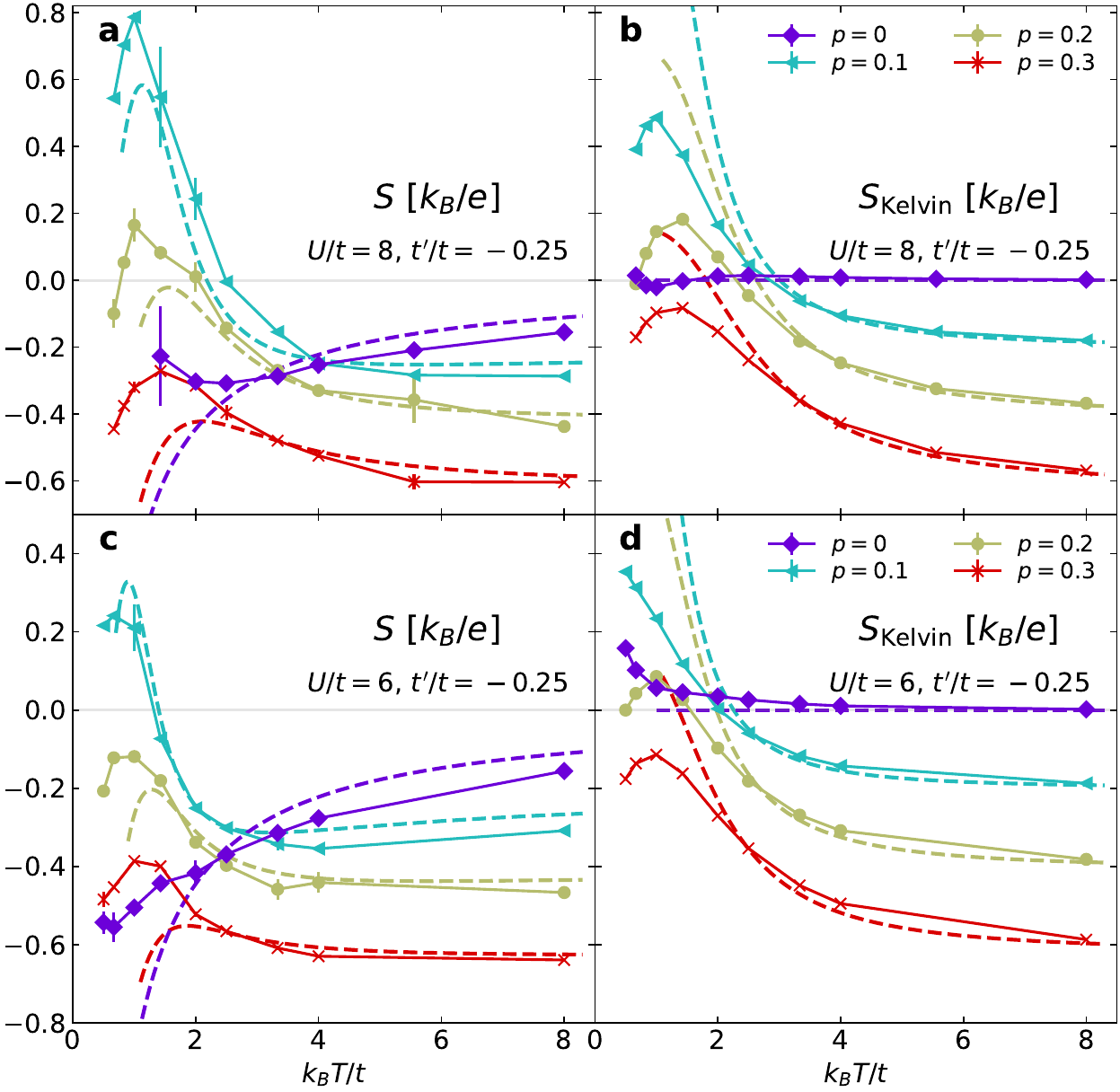}
    \caption{
    Hubbard model simulation data (solid lines, $S$ obtained with DQMC and MaxEnt, $S_\mathrm{Kelvin}$ obtained with DQMC) at high temperatures compared with corresponding atomic-limit results (dashed lines, from Supplementary Eqs.~\eqref{S_al} and \eqref{skelvin_al}), for $U/t=8$ in (\textbf{a}, \textbf{b}), and $U/t=6$ in (\textbf{c}, \textbf{d}). All panels have $t'/t=-0.25$.
 }
    \label{fig:al_tp}
\end{figure*}

\begin{figure*}
    \centering
    \includegraphics[width=0.8\textwidth]{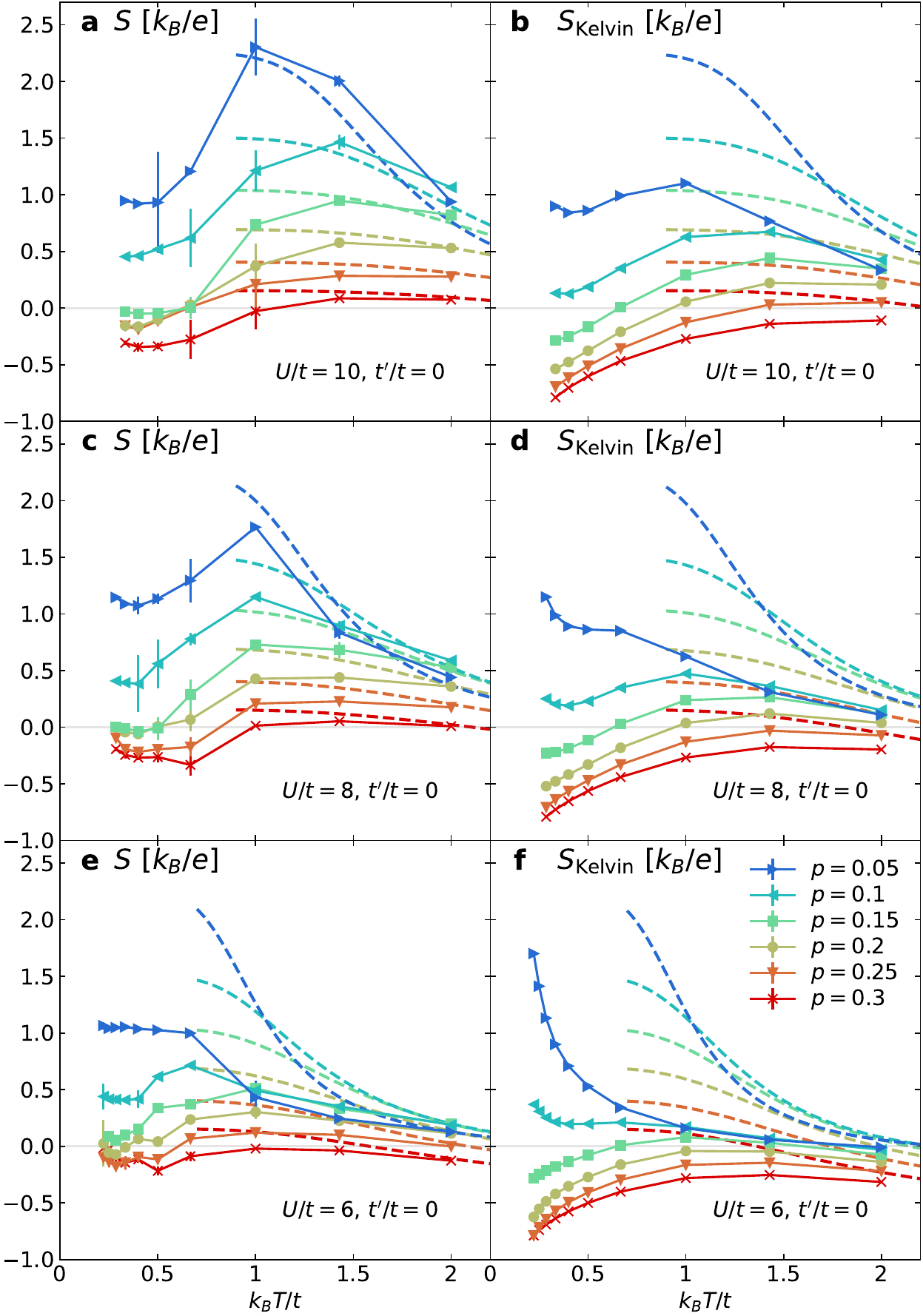}
    \caption{
Comparison similar to Supplementary Fig.~\ref{fig:al_tp} for $t'/t=0$, $T\lesssim 2$, (\textbf{a}, \textbf{b}) $U/t=10$, (\textbf{c}, \textbf{d}) $U/t=8$, and (\textbf{e}, \textbf{f}) $U/t=6$.
 }
    \label{fig:al_tp0}
\end{figure*}

In Supplementary Fig.~\ref{fig:atomiclimit_largeU}, we compare $S_\mathrm{Kelvin}$ calculated using DQMC with the atomic-limit approximation of $S_\mathrm{Kelvin}$, Supplementary Eq.~\eqref{skelvin_al}.
Large interactions $U/t=16$ and $U/t=12$ are selected.
At high temperatures, where the condition $t,t'\ll k_BT$ is satisfied, the simulation results match the atomic-limit approximations well.
As temperature decreases and this condition breaks down, $S_\mathrm{Kelvin}$ deviates from its atomic-limit approximation.

\begin{figure*}
    \centering
    \includegraphics[width=0.8\textwidth]{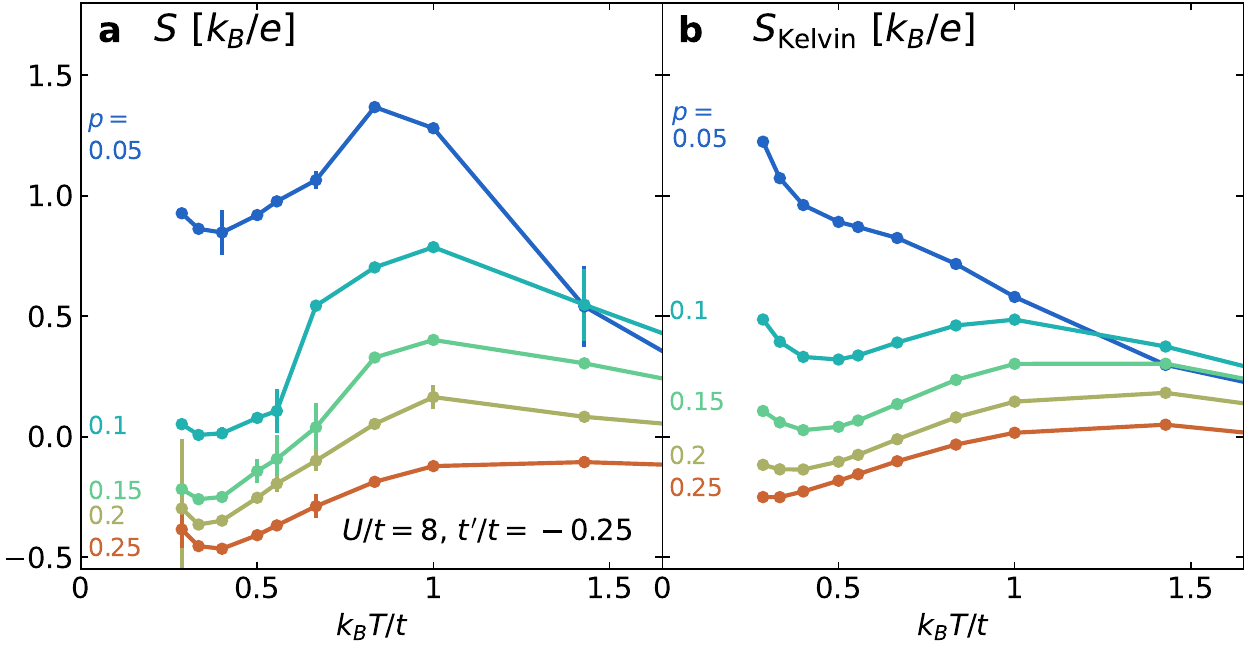}
    \caption{
    Temperature dependence of $S$ (\textbf{a}) and $S_\mathrm{Kelvin}$ (\textbf{b}), plotted in the same way as Fig.~3 in the main text, but for $U/t=8$ and $t'/t=-0.25$.
 }
    \label{fig:U8Tdep}
\end{figure*}

\begin{figure*}
    \centering
    \includegraphics[width=\textwidth]{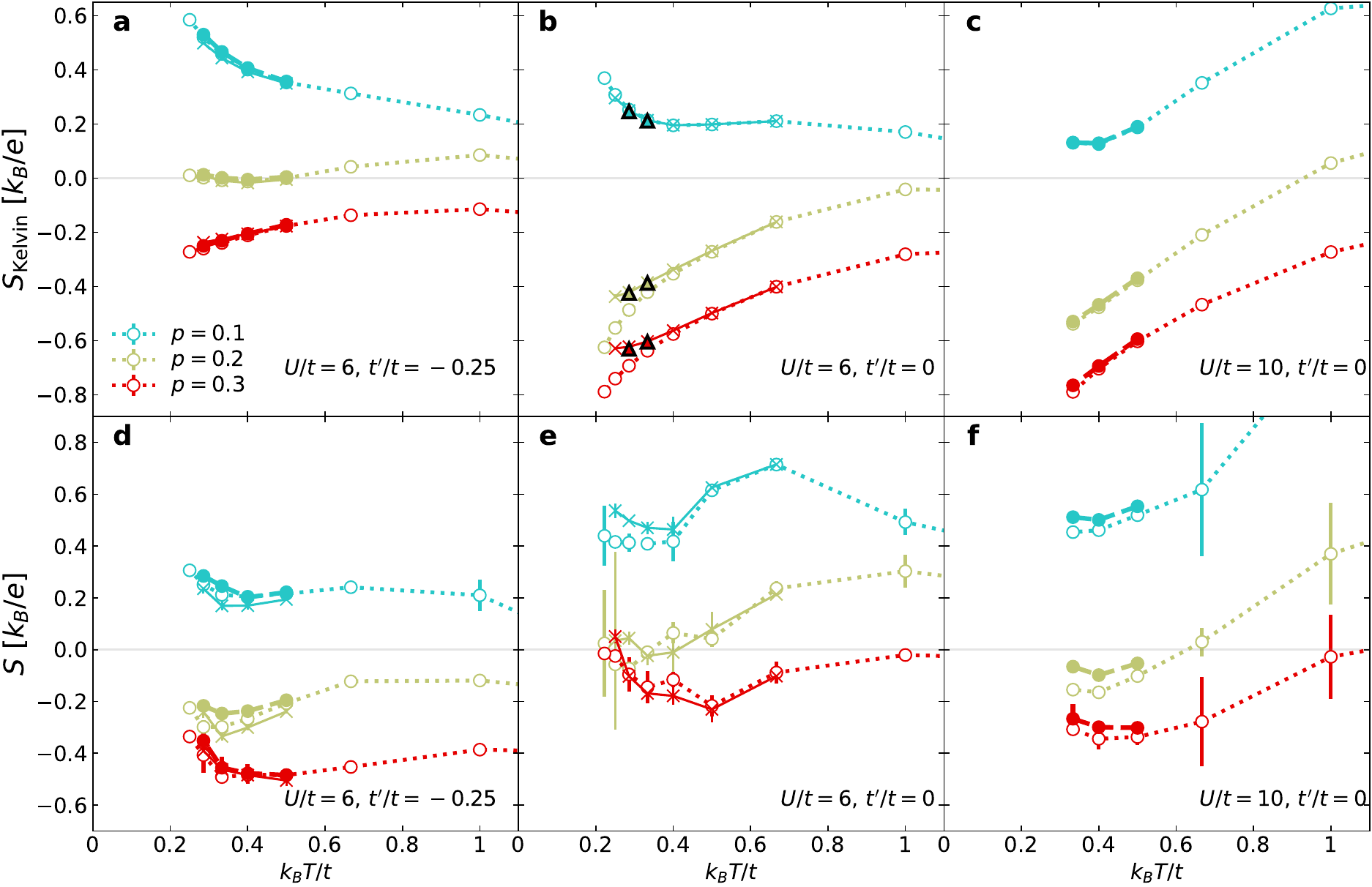}
    \caption{
    Finite size and Trotter error analysis for $S_\mathrm{Kelvin}$ (\textbf{a}, \textbf{b}, \textbf{c}) and $S$ (\textbf{d}, \textbf{e}, \textbf{f}).
    Dotted lines with open circles are obtained with $\dd\tau=0.05/t$ on an $8\times 8$ lattice, which are the parameters we use in the main text.
    Solid lines with crosses are obtained with $\dd\tau=0.05/t$ on a $12\times 12$ lattice.
Dashed lines with filled circles are obtained with $\dd\tau=0.025/t$ on an $8\times 8$ lattice.
 Triangles in (\textbf{b}) are obtained with $\dd\tau=0.05/t$ on a $16\times 16$ lattice.
For the solid and the dashed lines in (\textbf{d}), and the dashed lines in (\textbf{f}), the MaxEnt model functions for highest temperature ($\beta=2/t$) are chosen to be the spectral functions obtained with $\dd\tau=0.05/t$ on an $8\times 8$ lattice at $\beta=1.5/t$.
Similarly, for solid lines in (\textbf{e}), the MaxEnt model functions for highest temperature ($\beta=1.5/t$) are chosen to be the spectral functions obtained with $\dd\tau=0.05/t$ on an $8\times 8$ lattice at $\beta=1/t$.
 }
    \label{fig:finitesizeandtrotter}
\end{figure*}

Now, we derive the atomic-limit approximation for thermopower $S$.
Still using the occupation basis, and replacing $O_1$ and $O_2$ with $J_x$ or $J_{E,x}$ operators in Supplementary Eq.~\eqref{correlator_AL}, the $J_{E,x}-J_x$ and $J_x-J_x$ correlation functions to leading order are
\begin{align}
    &\frac{\ev{T_\tau J_{E,x}(\tau) J_x}}{N_xN_y} = \frac{\Tr (e^{-(\beta-\tau) (H_0-\mu N)} J_{E,x} e^{-\tau (H_0-\mu N)} J_x)}{N_xN_y\Tr e^{-\beta (H_0-\mu N)}}\\
 &= \frac{-16 t^2 t'-2U( t^2+2t'^2) }{Z_0^2} \times \nonumber \\
 &\quad \left( e^{3 \beta \mu+\frac{\beta U}{2}}  
+ e^{\beta\mu+\frac{\beta U}{2}} + e^{2\beta\mu+\tau U} +e^{2\beta\mu+\beta U-\tau U}  \right) \nonumber \\
    &+ \frac{2U( t^2+2t'^2)}{Z_0^2}\left( 2e^{3\beta\mu+\frac{\beta U}{2}}+e^{2\beta\mu+\tau U}+e^{2\beta\mu+\beta U-\tau U} \right) \label{jejcorr_al} \\
    &=\frac{\ev{T_\tau J_{x}(\tau) J_{E,x}}}{N_xN_y} \nonumber
\end{align}
and
\begin{align}
     &\frac{\ev{T_\tau J_x(\tau) J_x}}{N_xN_y}  = \frac{\Tr (e^{-(\beta-\tau) (H_0-\mu N)} J_x e^{-\tau (H_0-\mu N)} J_x)}{N_xN_y\Tr e^{-\beta (H_0-\mu N)}} \nonumber\\
&=\frac{4 (t^2+2t'^2)}{Z_0^2}\times  \nonumber \\
&\left( e^{3 \beta \mu+\frac{\beta U}{2}}  
+ e^{\beta\mu+\frac{\beta U}{2}} + e^{2\beta\mu+\tau U}+e^{2\beta\mu+\beta U-\tau U}  \right),\label{jjcorr_al}
\end{align}
where $Z_0=1+2e^{\beta U/2+\beta\mu}+e^{2\beta\mu}$.
Notice that any term of the form $ (e^{-(\beta-\tau)U}+e^{-\tau U})$ multiplied by a quantity independent of $\tau$ in $\ev{T_\tau O_1(\tau)O_2}+\ev{T_\tau O_2(\tau)O_1}$ corresponds to a delta function at $\omega=U$ in $\Re L_{O_1 O_2}(\omega) + \Re L_{O_2 O_1}(\omega)$ through Supplementary Eq.~\eqref{kernelO1O2}. Such terms do not contribute to the DC values of transport coefficients.
Any term independent of $\tau$ in $\ev{T_\tau O_1(\tau)O_2}+\ev{T_\tau O_2(\tau)O_1}$ corresponds to a delta function at $\omega=0$ in $\Re L_{O_1 O_2}(\omega) + \Re L_{O_2 O_1}(\omega)$. 
Summing up magnitudes of such terms provides the integrated weights of $\Re L_{O_1 O_2}(\omega) + \Re L_{O_2 O_1}(\omega)$ around $\omega=0$. So, using finite-frequency Onsager relations~\cite{Shastry_2008}, Supplementary Eqs.~\eqref{kernelO1O2}, \eqref{jejcorr_al}, and \eqref{jjcorr_al}, with $|\tilde{\epsilon}|<U$, we have
\begin{align}
        &\frac{1}{\pi}\int_{-|\tilde{\epsilon}|}^{+|\tilde{\epsilon}|}\Re L_{J_{E,x}J_x} (\omega) d\omega =\frac{1}{\pi}\int_{-|\tilde{\epsilon}|}^{+|\tilde{\epsilon}|}\Re L_{J_{x}J_{E,x}} (\omega) d\omega \nonumber  \\
        &= \frac{-16 t^2 t'-2U( t^2+2t'^2) }{Z_0^2}\left( e^{3 \beta \mu+\frac{\beta U}{2}}
+ e^{\beta\mu+\frac{\beta U}{2}}\right)\nonumber \\
    &\quad + \frac{4U( t^2+2t'^2)}{Z_0^2}e^{3\beta\mu+\frac{\beta U}{2}}, \label{jejcorr_alweight}
\end{align}
\begin{align}
     &\frac{1}{\pi}\int_{-|\tilde{\epsilon}|}^{+|\tilde{\epsilon}|} \Re L_{J_{x}J_x}  (\omega)  d\omega \nonumber \\
     &=\frac{4 (t^2+2t'^2)}{Z_0^2} \left( e^{3 \beta \mu+\frac{\beta U}{2}}  
+ e^{\beta\mu+\frac{\beta U}{2}}\right). \label{jjcorr_alweight}
\end{align}
Here, both $\Re L_{J_xJ_x}(\omega)$ and $\Re L_{J_{E,x}J_x}(\omega)$ are proportional to $\delta(\omega)$ at low frequencies, so they are both infinite at $\omega=0$.
To make both $\Re L_{J_xJ_x}(\omega=0)$ and $\Re L_{J_{E,x}J_x}(\omega=0)$ finite, we introduce a small scattering rate~\cite{mukerjee2007doping,PhysRevB.72.195109,PhysRevB.10.2186,PhysRevLett.122.186601}, or broadening effect, to both coefficients. The same scattering rate in both terms cancels out when we take their ratio and gives us the ratio of corresponding weights. 
Under this assumption, combining Supplementary Eqs.~\eqref{thermopower}, \eqref{jejcorr_alweight}, \eqref{jjcorr_alweight}, and that $\mathbf{J}_Q=\mathbf{J}_E-\mu\mathbf{J}$, we obtain the atomic-limit approximation of thermopower to leading order,
\begin{align}
    &S = \lim_{|\tilde{\epsilon}|\rightarrow 0} -\frac{\int_{-|\tilde{\epsilon}|}^{+|\tilde{\epsilon}|}d\omega (\Re L_{J_{E,x}J_x}(\omega)-\mu \Re L_{J_xJ_x}(\omega)) }{eT\int_{-|\tilde{\epsilon}|}^{+|\tilde{\epsilon}|}d\omega \Re L_{J_xJ_x}(\omega)}\nonumber \\
    & = \frac{4t^2 t'(e^{2 \beta \mu}  
+ 1)- U(t^2+2t'^2)e^{2\beta\mu}}{eT(t^2+2t'^2)(e^{2 \beta \mu}  
+ 1)} + \frac{\frac{U}{2}+\mu}{eT}. \label{S_al}
\end{align}

An interesting observation in the atomic limit is that $t$ and $t'$ affect $S$ (a transport property) in Supplementary Eq.~\eqref{S_al}, but not $S_\mathrm{Kelvin}$ (a thermodynamics property) in Supplementary Eq.~\eqref{skelvin_al}.
If we take $t'=0$, the expression Supplementary Eq.~\eqref{S_al} is equivalent to corresponding expressions of $S$ derived and discussed in Refs.~\cite{mukerjee2007doping,PhysRevB.72.195109,PhysRevB.10.2186,PhysRevLett.122.186601}, where the chemical potential is different from our definition by $U/2$ due to the difference in the Hamiltonian definition.

When we additionally impose the conditions $k_B T/U \ll 1$ (i.e. $\beta U \gg 1$) and $n<1$, and use $\mu$ as determined from Supplementary Eq.~\eqref{density_al} under these conditions, Supplementary Eq.~\eqref{skelvin_al} and the $t'=0$ case of Supplementary Eq.~\eqref{S_al} both approach the ``Heikes formula''~\cite{mukerjee2007doping,PhysRevB.72.195109,PhysRevB.10.2186,PhysRevB.13.647,PhysRevLett.122.186601}
\begin{align}
    S_\mathrm{Kelvin} = S = \frac{\frac{U}{2}+\mu}{eT} = \frac{k_B}{e} \ln \left[\frac{n}{2(1-n)}\right]. \label{heikes}
\end{align}
This ``Heikes limit'' in Supplementary Eq.~\eqref{heikes} produces a sign change at $p=1/3$~\cite{mukerjee2007doping,PhysRevB.72.195109,PhysRevB.10.2186,PhysRevB.13.647,phillips2009mottness,PhysRevB.82.214503,PhysRevLett.122.186601}.

In Supplementary Fig.~\ref{fig:al_tp}, we compare Hubbard model simulation results with the atomic limit of $S$ (Supplementary Eq.~\eqref{S_al}) and $S_\mathrm{Kelvin}$ (Supplementary Eq.~\eqref{skelvin_al}), for $t'/t=-0.25$ and $U/t = 6$ or $8$. 
The atomic-limit sign-change $p$ of $S$ shifts away from $1/3$ when $t'$ becomes non-zero, because of additional terms introduced by $t'$ in Supplementary Eq.~\eqref{S_al}.
Supplementary Figure~\ref{fig:al_tp0} presents the same comparison as Supplementary Fig.~\ref{fig:al_tp}, but for $t'/t=0$ and $U/t = 6$ to $10$ and focusing on lower temperatures.
 In all panels of Supplementary Figs.~\ref{fig:al_tp} and \ref{fig:al_tp0}, we see that simulation results (unsurprisingly) match the atomic-limit approximations at high temperatures but deviate as temperature decreases.

\section{Supplementary data}

For the sake of completeness, we show the temperature dependence of $S$ and $S_\mathrm{Kelvin}$ for $U/t=8$ and $t'/t=-0.25$ in Supplementary Fig.~\ref{fig:U8Tdep}. We find the behaviors of $S$ and $S_\mathrm{Kelvin}$ are qualitatively similar to the case of $U/t=6$ and $t'/t=-0.25$, shown in Fig.~3 in the main text. 

\section{Finite size and Trotter error}

We analyze finite-size effects and Trotter error for $S$ and $S_\mathrm{Kelvin}$ in Supplementary Fig.~\ref{fig:finitesizeandtrotter}.

Taking $U/t=6$ and $t'/t=-0.25$ as an example, differences between results obtained with $8\times 8$ and $12\times 12$ clusters are minimal for $S_\mathrm{Kelvin}$ in Supplementary Fig.~\ref{fig:finitesizeandtrotter}a, and are the same order of magnitude as the statistical errors for $S$ in Supplementary Fig.~\ref{fig:finitesizeandtrotter}d. 
The extent of finite-size effects changes with $t'$.
For $U/t=6$ and $t'/t=0$ in Supplementary Fig.~\ref{fig:finitesizeandtrotter}b, small finite-size discrepancies between $S_\mathrm{Kelvin}$ obtained with $8\times 8$ and $12\times 12$ clusters can be observed at high doping.
However, these differences do not impact the overall doping dependence.
Moreover, further increasing the lattice size to $16\times 16$ shows minimal difference compared to the $12\times 12$ lattice.
In Supplementary Fig.~\ref{fig:finitesizeandtrotter}e, differences between $S$ obtained with $8\times 8$ and $12\times 12$ clusters are the same order of magnitude as the statistical errors. 
Higher doping, smaller $U$, and lower temperature generally causes larger finite-size effects, as the system becomes more delocalized. Therefore,
our analysis up to $30\%$ doping, with $U/t=6$, including both $t'/t=-0.25$ and $t'/t=0$, and down to the lowest accessible temperatures provides an approximate upper limit for finite-size effects, given the parameters considered in this work. 

For two sets of parameters, $U/t=6$, $t'/t=-0.25$ and  $U/t=10$, $t'/t=0$, differences between results obtained with $\dd\tau=0.05/t$ and $\dd\tau=0.025/t$ are minimal for $S_\mathrm{Kelvin}$ in Supplementary Fig.~\ref{fig:finitesizeandtrotter}a and \ref{fig:finitesizeandtrotter}c, and are the same order of magnitude as the statistical errors of $S$ in Supplementary Fig.~\ref{fig:finitesizeandtrotter}d and \ref{fig:finitesizeandtrotter}f. 
Larger $U$ generally causes larger Trotter error, so our analysis up to $U/t=10$ provides an approximate upper limit for Trotter error for data presented in the main text of this work.

\end{document}